\documentclass{article}

\usepackage[nonatbib, final]{neurips_2021}
\usepackage[utf8]{inputenc} % allow utf-8 input
\usepackage[T1]{fontenc}    % use 8-bit T1 fonts
\usepackage{hyperref}       % hyperlinks
\usepackage{url}            % simple URL typesetting
\usepackage{booktabs}       % professional-quality tables
\usepackage{amsfonts}       % blackboard math symbols
\usepackage{nicefrac}       % compact symbols for 1/2, etc.
\usepackage{microtype}      % microtypography
\usepackage{xcolor}         % colors
\usepackage{graphicx}
\usepackage{cite}

\usepackage{wrapfig}
\usepackage{capt-of}
\usepackage{caption}
\usepackage{comment}

\title{Data-Driven Computational Imaging for \\ Scientific Discovery}

\author{Andrew Olsen\thanks{
Equal contribution. Correspondence to
\texttt{vidyag@berkeley.edu}.
}~ $^1$, Yolanda Hu$^{*1}$, Vidya Ganapati$^{1,2}$\\
    $^1$ Swarthmore College, $^2$ Lawrence Berkeley National Laboratory
  }

\begin{document}

\maketitle

\begin{abstract}

In computational imaging, hardware for signal sampling and software for object reconstruction are designed in tandem for improved capability. Examples of such systems include computed tomography (CT), magnetic resonance imaging (MRI), and superresolution microscopy. In contrast to more traditional cameras, in these devices, indirect measurements are taken and computational algorithms are used for reconstruction. This allows for advanced capabilities such as super-resolution or 3-dimensional imaging, pushing forward the frontier of scientific discovery. However, these techniques generally require a large number of measurements, causing low throughput, motion artifacts, and/or radiation damage, limiting applications. Data-driven approaches to reducing the number of measurements needed have been proposed, but they predominately require a ground truth or reference dataset, which may be impossible to collect. This work outlines a self-supervised approach and explores the future work that is necessary to make such a technique usable for real applications. Light-emitting diode (LED) array microscopy, a modality that allows visualization of transparent objects in two and three dimensions with high resolution and field-of-view, is used as an illustrative example. We release our code at \url{https://github.com/vganapati/LED_PVAE} and our experimental data at \url{https://doi.org/10.6084/m9.figshare.21232088}.

\end{abstract}

\begin{comment}

In computational imaging, hardware for signal sampling and software for object reconstruction are designed in tandem for improved capability. Examples of such systems include computed tomography (CT), magnetic resonance imaging (MRI), and superresolution microscopy. In contrast to more traditional cameras, in these devices, indirect measurements are taken and computational algorithms are used for reconstruction. This allows for advanced capabilities such as super-resolution or 3-dimensional imaging, pushing forward the frontier of scientific discovery. However, these techniques generally require a large number of measurements, causing low throughput, motion artifacts, and/or radiation damage, limiting applications. Data-driven approaches to reducing the number of measurements needed have been proposed, but they predominately require a ground truth or reference dataset, which may be impossible to collect. This work outlines a self-supervised approach and explores the future work that is necessary to make such a technique usable for real applications. Light-emitting diode (LED) array microscopy, a modality that allows visualization of transparent objects in two and three dimensions with high resolution and field-of-view, is used as an illustrative example. We release our code at https://github.com/vganapati/LED_PVAE and our experimental data at https://doi.org/10.6084/m9.figshare.21232088.

\end{comment}

\section{Introduction}

Computational imaging systems, which reconstruct objects from indirect measurements, are ubiquitous in modern scientific research. For example, computed tomography allows for 3-dimensional visualization by collecting x-ray projections through an object, and is used in a range of fields including medicine \cite{rubin_computed_2014}, biology \cite{wise_micro-computed_2013}, materials science \cite{garcea_x-ray_2018}, and geoscience \cite{cnudde_high-resolution_2013}. Another computational imaging modality, light-emitting diode (LED) array microscopy (also known as Fourier ptychographic microscopy), has shown success in quantitative 2-dimensional and 3-dimensional phase imaging \cite{zheng_wide-field_2013, ou_quantitative_2013, tian_multiplexed_2014, tian_computational_2015, tian_3d_2015, horstmeyer_diffraction_2016,li_high-speed_2019}, with applications in pathology \cite{majeed_quantitative_2017, jin_tomographic_2017} and biology \cite{nandakumar_isotropic_2012,sandoz_label_2018}.

Though computational imaging methods have achieved a high degree of success in spatial resolution, the temporal resolution (imaging speed) remains low. Increasing speed remains an active area of research, as success will allow visualization in unprecedented regimes of spatial and temporal resolution. In x-ray computed tomography, for example, thousands of 2-dimensional images must be collected for eventual object reconstruction. In LED array microscopy, the light source of a conventional wide-field microscope is replaced with a 2-dimensional LED array. Each LED of the array is individually addressable with tunable brightness, allowing different patterns to be illuminated. Generally, one image is collected per LED, and arrays may consist of hundreds of LEDs. 

Data-driven deep learning methods, using a reference training dataset of measurement-reconstruction pairs, have been widely proposed to improve the temporal resolution of computational imaging. However, the necessity of a training dataset makes the technique prohibitive in many applications of scientific discovery. A chicken-and-egg problem arises in the case of fragile or live specimens: without a reference object dataset, we cannot create a faster imaging method, but without the faster imaging method the training object dataset cannot be obtained. In this work, we outline a reconstruction method that \textit{only} requires a representative dataset of sparse or partial measurements on each object. To circumvent the need for complete training dataset pairs, we look to jointly reconstruct a set of similar objects, each with a low number of measurements. By pooling information from measurements across the set and incorporating the known forward physics of imaging, we aim to jointly infer the prior distribution and posterior distributions. We aim to allow for improved reconstructions with fewer measurements per object by using information from multiple similar objects.

More precisely, computational imaging aims to reconstruct some object $O$ from a sequence of $n$ noisy measurements $M = [M_1, M_2, ..., M_n]$. We aim to lower the total number of measurements $n$ to minimize data acquisition time. We assume that we have a set of $m$ objects $\{O_1, O_2, ..., O_m\}$, sampled from some distribution $P(O)$, and we aim to reconstruct all objects in the set. For each of the $m$ objects, we are allowed $n$ measurements. Each sequence of measurements for an object~$j$, $M_j = [M_{j1}, M_{j2}, ..., M_{jn}]$ is obtained with chosen hardware parameters $p_j = [p_{j1}, p_{j2}, ..., p_{jn}]$ (e.g. rotation angles in the case of computed tomography or the LED illumination patterns in the case of LED array microscopy). We assume that the forward model physics $P(M | O; p) = P(M | O)$ is known. For every object $O$, we aim to find the posterior distribution $P(O | M) = \frac{P(M | O) P(O)}{P(M)}$. The following problems arise in finding the posterior: (1) construction of the prior $P(O)$ with no directly observed $O$, only indirect measurements $M$ on each object of the set, and (2) calculating $P(O | M)$ in a tractable manner. To efficiently solve this problem, we create a novel technique through a reformulation of variational autoencoders. The probabilistic formulation considered in this work permits uncertainty quantification, in contrast to most reconstruction algorithms that only yield a point estimate.

\section{Related Work}

Deep learning has been widely applied to reduce the data acquisition burden of computational imaging systems. In one line of research, training pairs of sparse measurements and corresponding high quality reconstructions are used to train a deep convolutional neural network, and implicitly embed prior information \cite{kamilov_learning_2015, kalantari_learning-based_2016, sinha_solving_2016, kappeler_ptychnet_2017, li_imaging_2017, mardani_recurrent_2017, chen_low-dose_2017, gul_spatial_2017, jin_deep_2017, shimobaba_computational_2017, sinha_lensless_2017, mccann_review_2017, rivenson_deep_2017-1, rivenson_toward_2018, borhani_learning_2018,  goy_low_2018, nguyen_2d_2018, sun_efficient_2018, wu_extended_2018-1, nguyen_deep_2018, boominathan_phase_2018, lucas_using_2018, schwartz_deepisp_2018, DiSpirito_reconstructing_2020, li_fast_2020}. Subsequent sparse measurements can be reconstructed with a forward pass of the trained neural network, with the benefit of avoiding computationally costly iterative algorithms.

Deep neural network approaches for object reconstruction have the advantage of incorporating knowledge about prior data and fast inference, but more traditional iterative (model-based) methods have the advantage of utilizing the known forward physics model (i.e.\ how measurements are generated, given the object). The advantages of these two approaches are combined by unrolling an iterative method, with each iteration forming a layer of a neural network \cite{gregor_fast_2010, andrychowicz_learning_2016, adler_solving_2017, zhang_learning_2017, aggarwal_modl_2017, bostan_learning-based_2018, hammernik_learning_2018, li_deep_2019, monga_algorithm_2019, diamond_unrolled_2018, Bo_FompNet_2017, nakarmi_multi-scale_2020, wu_simba_2020}. This unrolled deep neural network can be trained to optimize iterative algorithm hyperparameters for a given training dataset, effectively optimizing an optimizer. The unrolled iterative methods have been shown to require less training data and time than a convolutional neural network approach, due to the incorporation of the forward model. 

Building off of this literature, a second body of approaches attempts to include the measurement process in neural network training, to discern the optimal measurement parameters (e.g. the LED illumination patterns in LED array microscopy) for sparse sampling and subsequent reconstruction. In these works, high quality reconstructions are needed, and corresponding noisy measurements are emulated with the known forward physics. The measurement process is included as the encoder part of an autoencoder neural network, and co-optimized with the reconstruction algorithm, which forms the decoder.  Many works use a convolutional neural network as the decoder \cite{adler_deep_2016, Iliadis_DeepBinaryMask_2016, adler_block-based_2017, xie_adaptive_2017, mousavi_DeepCodec_2017, lohit_convolutional_2018, higham_deep_2018, du_fully_2017, grover_uncertainty_2019, qiao_deep_2020, bahadir_deep-learning-based_2020,chakrabarti_learning_2016, robey_optimal_2018, haim_depth_2018, elmalem_learned_2018, diederich_using_2018, cheng_illumination_2019, cheng_deep_2019, muthumbi_learned_2019, hershko_multicolor_2019, hougne_learned_2020, Nehme_DeepSTORM3D_2020} and others use an unrolled iterative solver \cite{sitzmann_end_2018,kellman_physics-based_2019, kellman_data-driven_2019}. Co-optimizing the measurement parameters has the benefit of reducing the measurements required for computational imaging further than keeping them fixed during training. However, this approach still requires a reference training dataset.

In this work, we look to remove the need for ground-truth or reference reconstructions. We aim to create a reconstruction method that \textit{only} requires a representative dataset of sparse measurements on each object. This task has been previously undertaken, usually with generative adversarial networks \cite{kabkab_task-aware_2018, bora2018ambientgan, kuanar_low_2019, Cole_2021_ICCV, xu_fast_2019, tamir_unsupervised_2020, gan_image_2020, Zhussip_2019_CVPR, liu_rare_2020}. The intuition is that by using different experimental measurement parameters for every object of the set, it is possible to build a general understanding of what an object should look like (i.e. the prior). However, these methods all lack probabilistic outputs. In this work, we propose a method based on variational autoencoders that solves for the posterior distribution in a principled manner and outline some of the progress required to make this technique usable for scientific discovery.

%1. Deep neural networks to replace the reconstruction algorithm
%Need full training pairs
%High burden on training
%Not generalizable
%
%2. Optimizing optimization hyperparameters with training pairs (unrolled networks)
%need full training pairs
%less parameters to train
%more generalizable
%less data needed to train
%
%3. co-optimization of hardware parameters with the neural network
%need full training pairs
%both with unrolled networks and with full deep neural networks
%
%4. Data-Driven Computational Imaging for Scientific Discovery
%don't have full training pairs of measurements - reconstructions
%live samples
%fragile samples

\section{Physics-Informed Variational Autoencoder}

We assume that we have a set of $m$ objects $\{O_1, O_2, ..., O_m\}$ drawn from $P(O)$, where $P(O)$ is unknown and we cannot directly measure $O$. For each object $O$, we are allowed to take $n$ indirect measurements $M = [M_1, M_2, ..., M_n]$. The measurement $M$ on an object $O$ is obtained with chosen hardware parameters $p = [p_1, p_2, ..., p_n]$, and the forward model $P(M | O; p) = P(M | O)$ is known through the physics of image formation. We aim to determine the posterior distribution $P(O | M)$ for every object in the set.

\begin{wrapfigure}{l}{0.55\textwidth}
  \vspace{-1.5em}
  \begin{center}
    \includegraphics[width=0.55\textwidth]{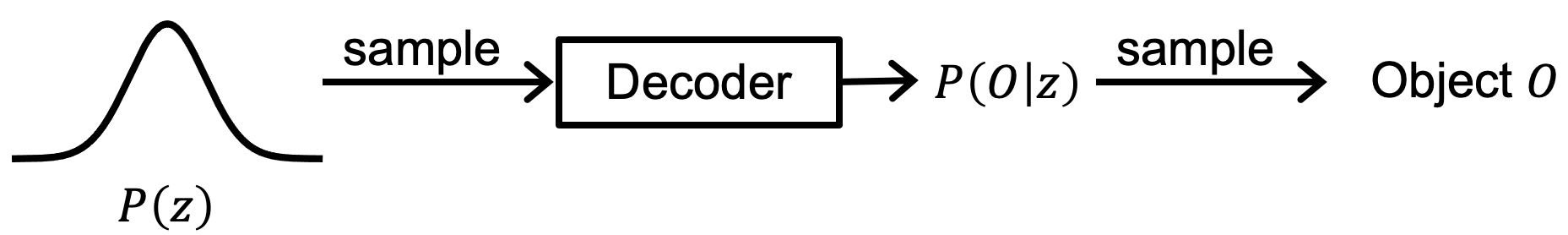}
  \end{center}
  \caption{Generating object $O$ from a latent variable $z$.}
  \label{fig:gen}
  \vspace{0em}
\end{wrapfigure}

We propose a general framework for posterior estimation that is inspired by the mathematics of the variational autoencoder \cite{kingma2014autoencoding, doersch2021tutorial}. In a variational autoencoder, the goal is to learn how to generate new examples, sampled from the same underlying probability distribution as a training dataset of objects. To accomplish this task, a latent random variable $z$ is created that describes the space on a lower-dimensional manifold. A deep neural network defines a function (the ``decoder'') from a sample of $z$ to a probability distribution $P(O | z)$, see Fig.~\ref{fig:gen}. Deep neural networks are chosen for function approximation due to their theoretical ability to approximate any function \cite{cybenko_approximation_1989, hornik_approximation_1991, leshno_multilayer_1993} and practical success in approximating high-dimensional functions \cite{lecun_deep_2015}. The parameters of the deep neural network are optimized to maximize the probability of generating the objects in the training set.

In this work, we aim to find the posterior probability distribution $P(O | M)$, where $O$ is the object being reconstructed and $M$ is the measurement. In our case, we only have a dataset of noisy measurements $M$ and no ground truth objects $O$, but a known forward model, $P(M | O)$. Thus, instead of maximizing the probability of generating $O$, we can maximize the probability of generating $M$, a formulation we call the ``physics-informed variational autoencoder.''

\begin{figure}[h]
  \vspace{-1.0em}
\centering
\includegraphics[width=1\textwidth]{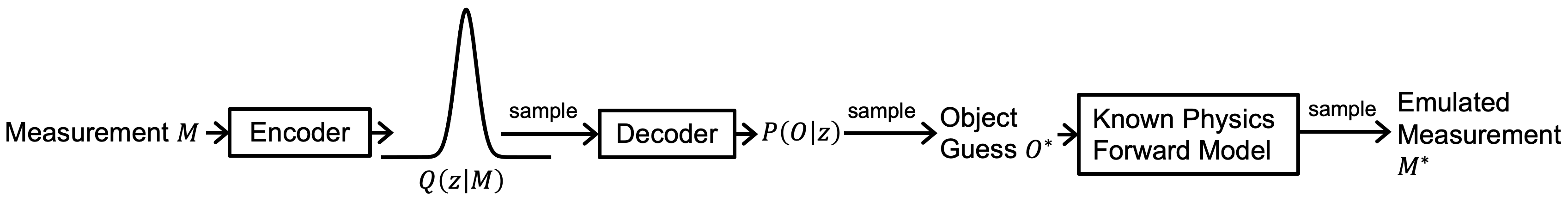}
\caption{The physics-informed variational autoencoder.}
\label{fig:vae}
  \vspace{-0.5em}
\end{figure}

We aim to maximize $P(M)= \int \int P(M|O)P(O|z)P(z) dO dz$. To compute this integral in a computationally tractable manner, we can approximate with sampled values. However, for most values of $z$ and $O$, the probability $P(M|O,z)$ is close to zero, causing poor scaling of sampled estimates. Similar to a variational autoencoder, our framework solves this problem by estimating the parameters of $P(z|M)$ by processing the measurements $M$ using a function with trainable parameters (called the ``encoder,'' usually a deep neural network). The estimate of $P(z|M)$ is denoted $Q(z|M)$. The Kullback–Leibler divergence between the distributions is given by $D[Q(z|M) || P(z|M)] = E_{z \sim Q}[\log Q(z|M) - \log P(z|M)]$. We also have, by Bayes' Theorem, $\log P(z|M) = \log P(M|z) + \log P(z) - \log P(M)$. Combining the expressions yields:

\[
 \log P(M) - D[Q(z|M) || P(z|M)] = E_{z \sim Q} \left[ \int P(M|O)P(O|z)dO\right] - D\left[\log Q(z|M) || \log P(z) \right]. 
\]

The first term on the right side of this expression can be estimated with sampled values. As Kullback–Leibler divergence is always $\geq 0$ and reaches $0$ when $Q(z|M) = P(z|M)$, maximizing the right side (defined here as the loss) during training causes $P(M)$ to be maximized while forcing $Q(z|M)$ towards $P(z|M)$. In contrast to a conventional variational autoencoder, we do \textit{not} attempt to use this formulation to synthesize arbitrary objects $O$ by sampling $P(z)$ directly. This framework only attempts reconstruction on the training examples themselves. After training, $P(O|M)$ for every object can be sampled by first sampling $Q(z|M)$ then $P(O|z)$, see Fig.~\ref{fig:vae}. Crucially, unlike most data-driven approaches to reconstruction in computational imaging, this framework assumes that no ground truth dataset of objects $O$ is available.

\section{Light-Emitting Diode (LED) Array Microscopy}

\begin{wrapfigure}{l}{0.2\textwidth}
  \vspace{-1.5em}
  \begin{center}
    \includegraphics[width=0.2\textwidth]{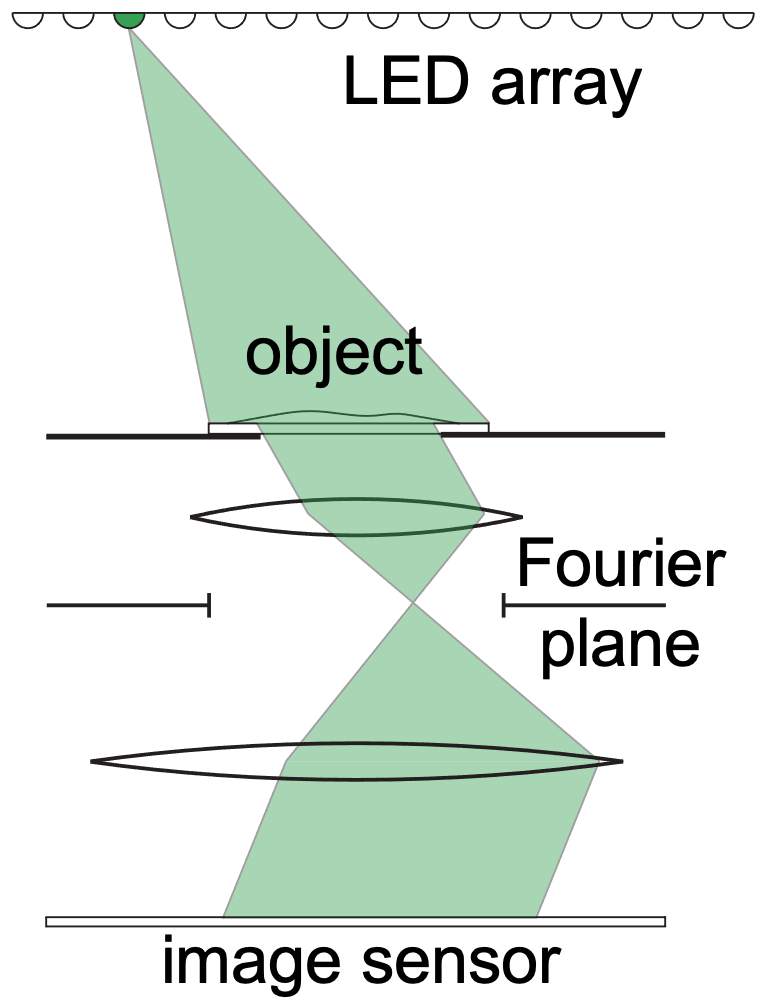}
  \end{center}
  \caption{Schematic of the LED array microscope.}
  \label{fig:FPM}
  \vspace{-2em}
\end{wrapfigure}

%
%\begin{figure}[h]
%\centering
%\includegraphics[width=0.2\textwidth]{figures/fpm.png}
%\caption{Schematic of the LED array microscope, a computational imaging system.}
%\label{fig:FPM}
%\end{figure}

We prototype the physics-informed variational autoencoder with LED array microscopy as the example system. An LED array microscope is a device created by replacing the illumination source of a standard wide-field microscope with a 2-dimensional LED array \cite{zheng_wide-field_2013}, see Fig.~\ref{fig:FPM}. This device has great potential for enabling scientific advances in biology due to its ability to quantitatively measure unstained samples with a simple hardware platform \cite{konda_fourier_2020}. 

Every LED of the microscope's array is individually addressable with tunable brightness, allowing different patterns to be illuminated. For every object (e.g.\ specimen or field of view) imaged, a set of $n$ illumination patterns with parameters $p = [p_1, p_2, ..., p_n]$ are chosen, resulting in an image stack $M = [M_1, M_2, ..., M_n]$. The resulting measurements $M$ are conventionally post-processed with an iterative optimization algorithm to yield a 2- or 3-dimensional amplitude and phase reconstruction of the object $O$. We aim to increase the temporal resolution (i.e.\ reduce acquisition time) by reducing the number of images $n$ needed per object. Results from LED array microscopy can be translated to other computational imaging modalities such as computed tomography by changing the forward physics model $P(M|O; p)$ in Fig.~\ref{fig:vae}.

\subsection{Neural Network Architecture}

\begin{figure}[h]
  \vspace{0em}
\centering
\includegraphics[width=1.0\textwidth]{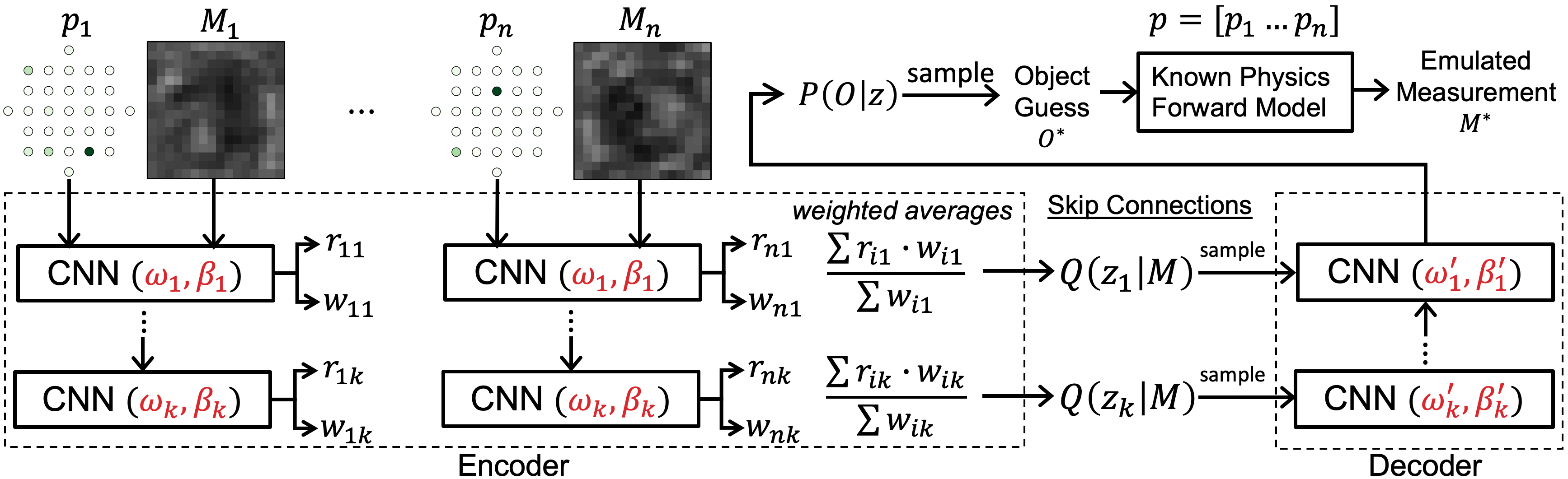}
\caption{Permutation-invariant network architecture with $k$ encoding and decoding layers.}
\label{fig:perm_invar}
  \vspace{-1em}
\end{figure}

The overall deep neural network design for the physics-informed variational autoencoder is shown in Fig.~\ref{fig:perm_invar}. A permutation-invariant design is created so that the reconstruction is agnostic to the order of the $n$ input measurements. The network takes the same basic shape as a U-Net, a network design that has shown success in image processing tasks such as segmentation \cite{ronneberger_u-net_2015} and super-resolution \cite{hu_runet_2019}. Each measurement parameter $p_i$ and corresponding measurement $M_i$ (in the case of LED array microscopy, the illumination pattern used and the corresponding intensity image collected, respectively) are concatenated and used as input to this architecture. The skip connections are parametrized as Gaussian distributions with mean and variance determined through a weighted average from all inputs $i=1,2,...,n$.

\subsection{Datasets}

\subsubsection{Synthetic Foam}
\label{sec:dataset}

\begin{wrapfigure}{r}{0.6\textwidth}
  \vspace{-2em}
  \begin{center}
    \includegraphics[width=0.6\textwidth]{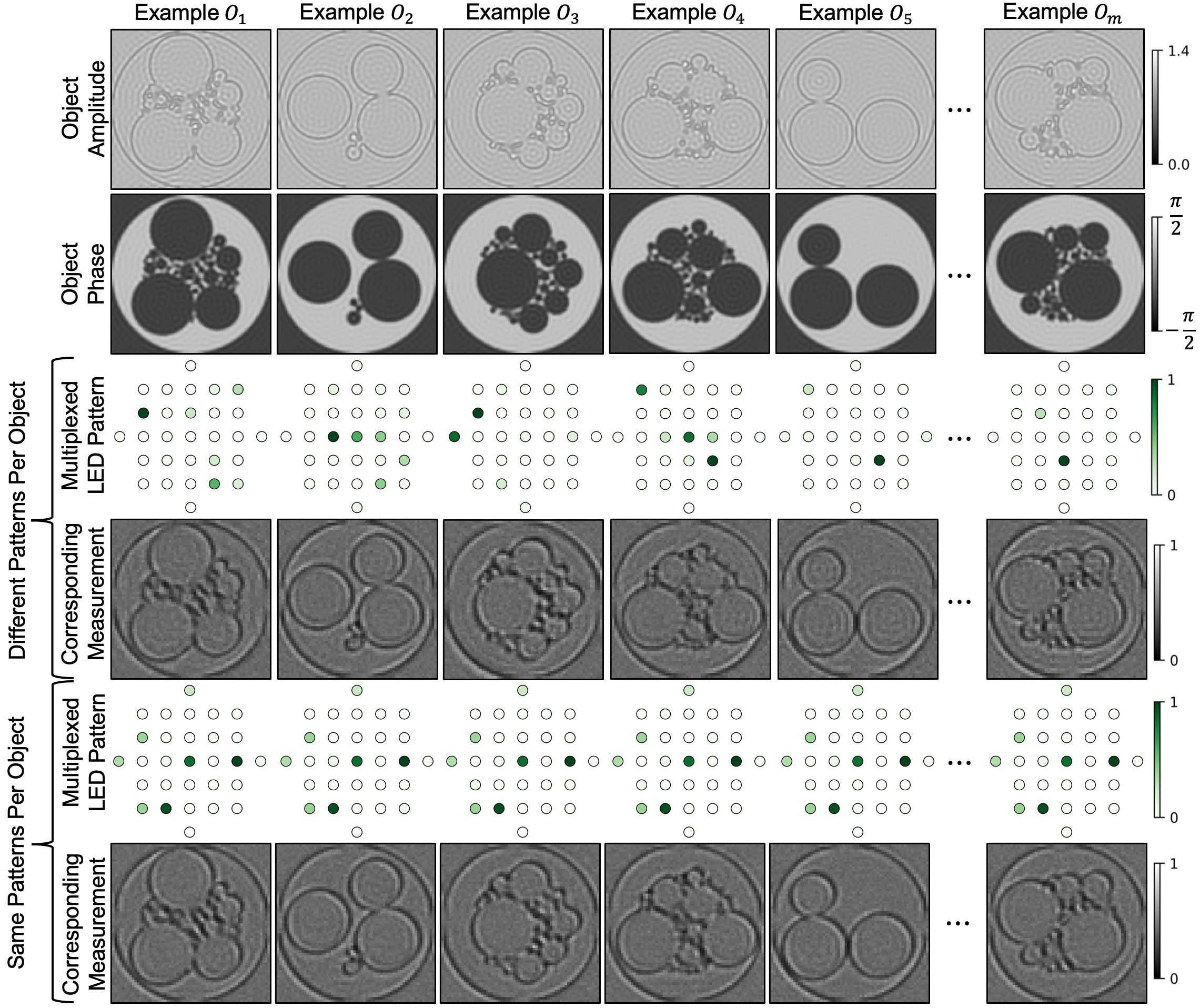}
  \end{center}
  \caption{Example objects from the synthetic ``foam'' dataset. We emulate measurements with different illumination patterns for each object (middle two rows) and the same illumination pattern for every object (bottom two rows).}
  \label{fig:dataset}
  \vspace{-1.0em}
\end{wrapfigure}

A synthetic dataset is created from foam images generated through the Python package \textsc{XDesign} \cite{noauthor_xdesign_nodate}, and simulated as 2D complex objects, see Fig.~\ref{fig:dataset}. The measurement acquisition procedure is simulated using the known forward physics of LED array microscopy and statistics of Poisson noise, as in \cite{robey_optimal_2018}. Each object of the dataset is assumed to be measured using a single random illumination pattern on a 2D array of 29 LEDs.

We model the distribution of possible multiplexed illumination patterns $P(p)$ with a multivariate Dirichlet distribution. We sample this distribution $n$ times for every object, where $n=1$ for this synthetic dataset. The Dirichlet distribution is defined over $x_1, \dots , x_l$ where $ \sum x_i = 1$. Here, $l$ is the number of LEDs, with each LED $i$ having brightness $x_i$. The constraint of $ \sum x_i = 1$ prevents detector saturation if the exposure is set judiciously. The Dirichlet distribution is parameterized by $\alpha_1, ..., \alpha_l$, with $\alpha_i > 0$. If all $\alpha_i = 1$, then all patterns are equally likely. In this synthetic dataset, all $\alpha_i = 0.1$, favoring sparser patterns. To probe the impact of measurement diversity, we create a corresponding set of measurements where a single illumination pattern is sampled from the Dirichlet distribution and used to measure every object (i.e. $P(p)$ is deterministic), see Fig.~\ref{fig:dataset}.

\begin{wrapfigure}{r}{0.6\textwidth}
  \vspace{-3em}
  \begin{center}
    \includegraphics[width=0.6\textwidth]{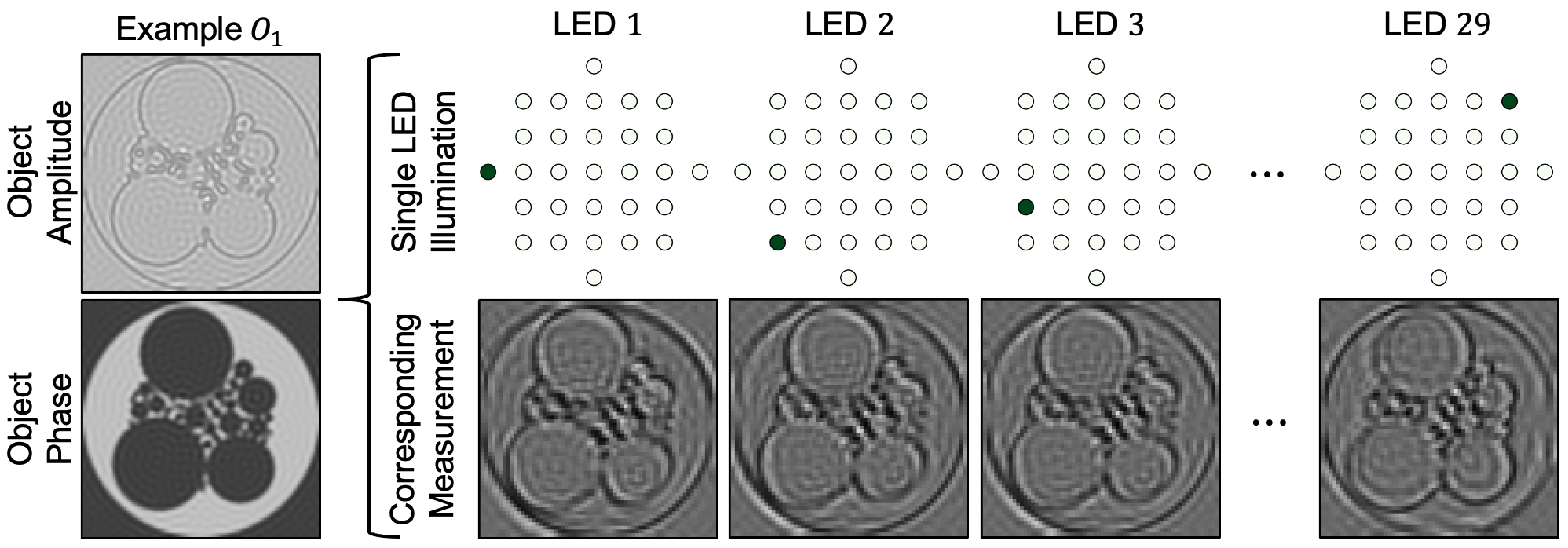}
  \end{center}
  \vspace{-1em}
  \caption{Sequential single LED illumination.}
  \label{fig:single_led}
  \vspace{-3em}
\end{wrapfigure}

In standard LED array microscopy, the LEDs are illuminated sequentially, one at a time. This entire stack of images is post-processed with an iterative algorithm for the final reconstruction. This image stack is visualized in Fig.~\ref{fig:single_led}.

\subsubsection{Synthetic 3-Dimensional MNIST}

\begin{figure}[h]
\vspace{-1.0em}
\centering
\includegraphics[width=1\textwidth]{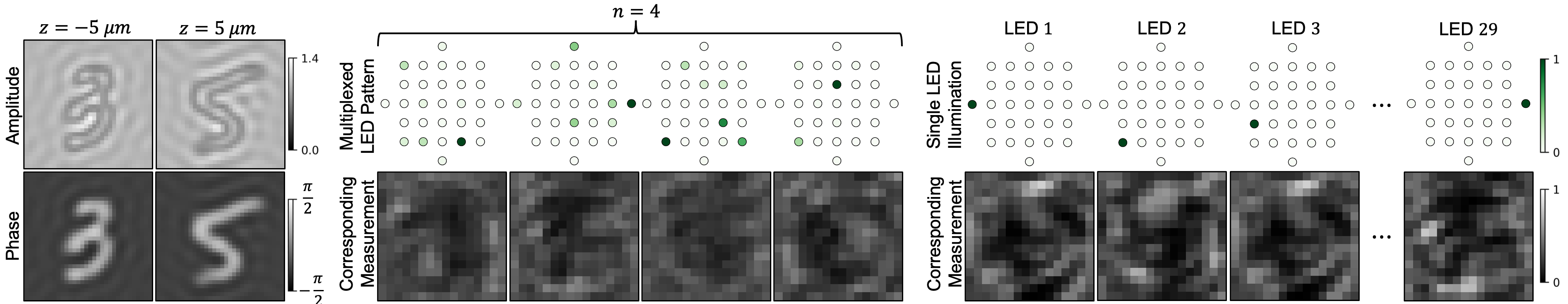}
\caption{Example object from the synthetic 3-dimensional MNIST dataset.}
\label{fig:dataset_mnist}
\vspace{-0.5em}
\end{figure}

A second synthetic dataset is created utilizing MNIST digits \cite{lecun-mnisthandwrittendigit-2010}. Each example consists of two MNIST digits in axial planes $10$ $\mu$m apart. Each of the two MNIST digits is approximated as a thin phase object and the resulting intensity image is modeled using the assumptions of \cite{tian_3d_2015}. An example object is shown in Fig.~\ref{fig:dataset_mnist}.

\subsubsection{Experimental Frog Blood Smears}

\begin{figure}
    \centering
    \begin{minipage}{0.51\textwidth}
        \centering
        \includegraphics[width=1.0\textwidth]{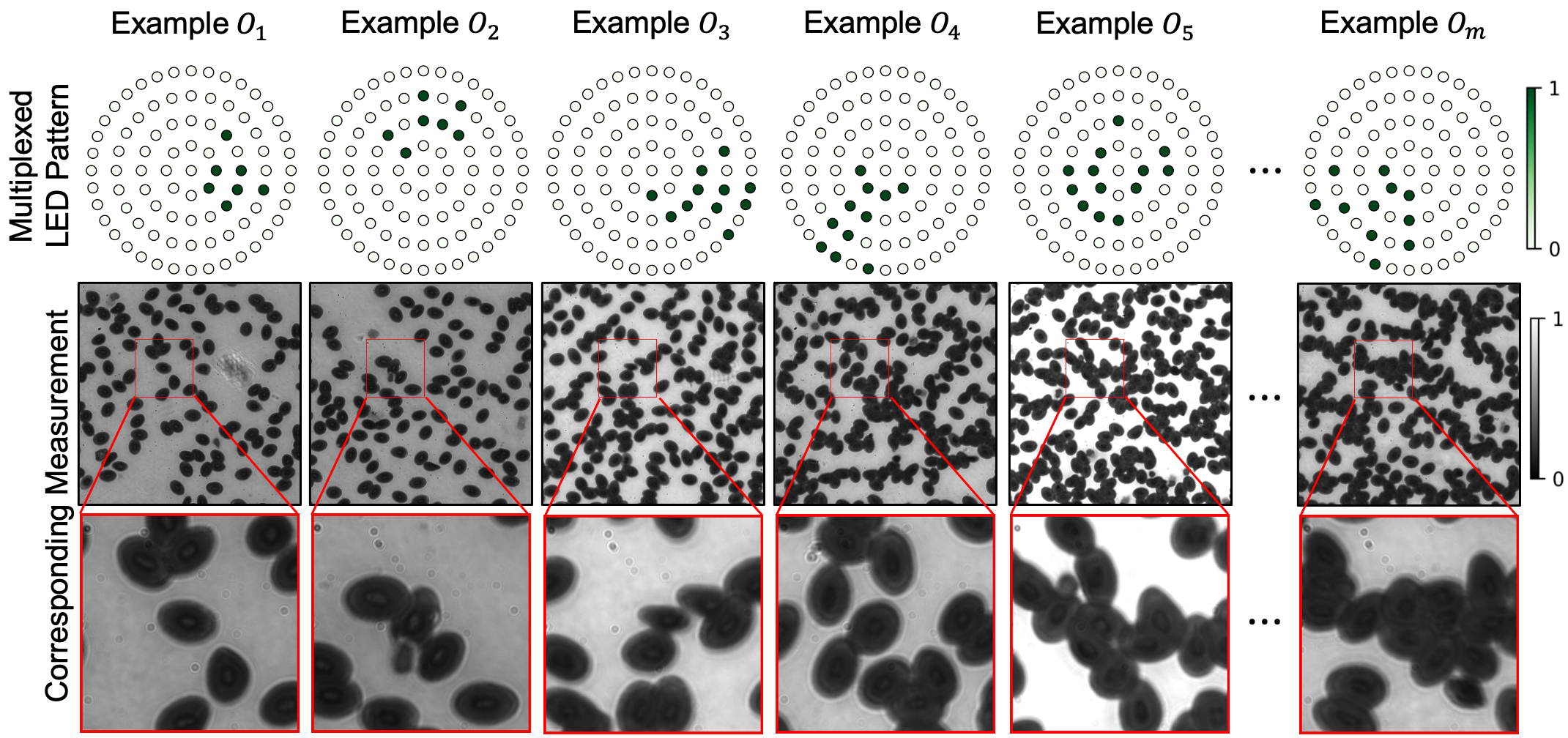} % first figure itself
        \caption{Experimental dataset collected.}
        \label{fig:dataset_real}
    \end{minipage}\hfill
    \begin{minipage}{0.45\textwidth}
        \centering
        \includegraphics[width=1.0\textwidth]{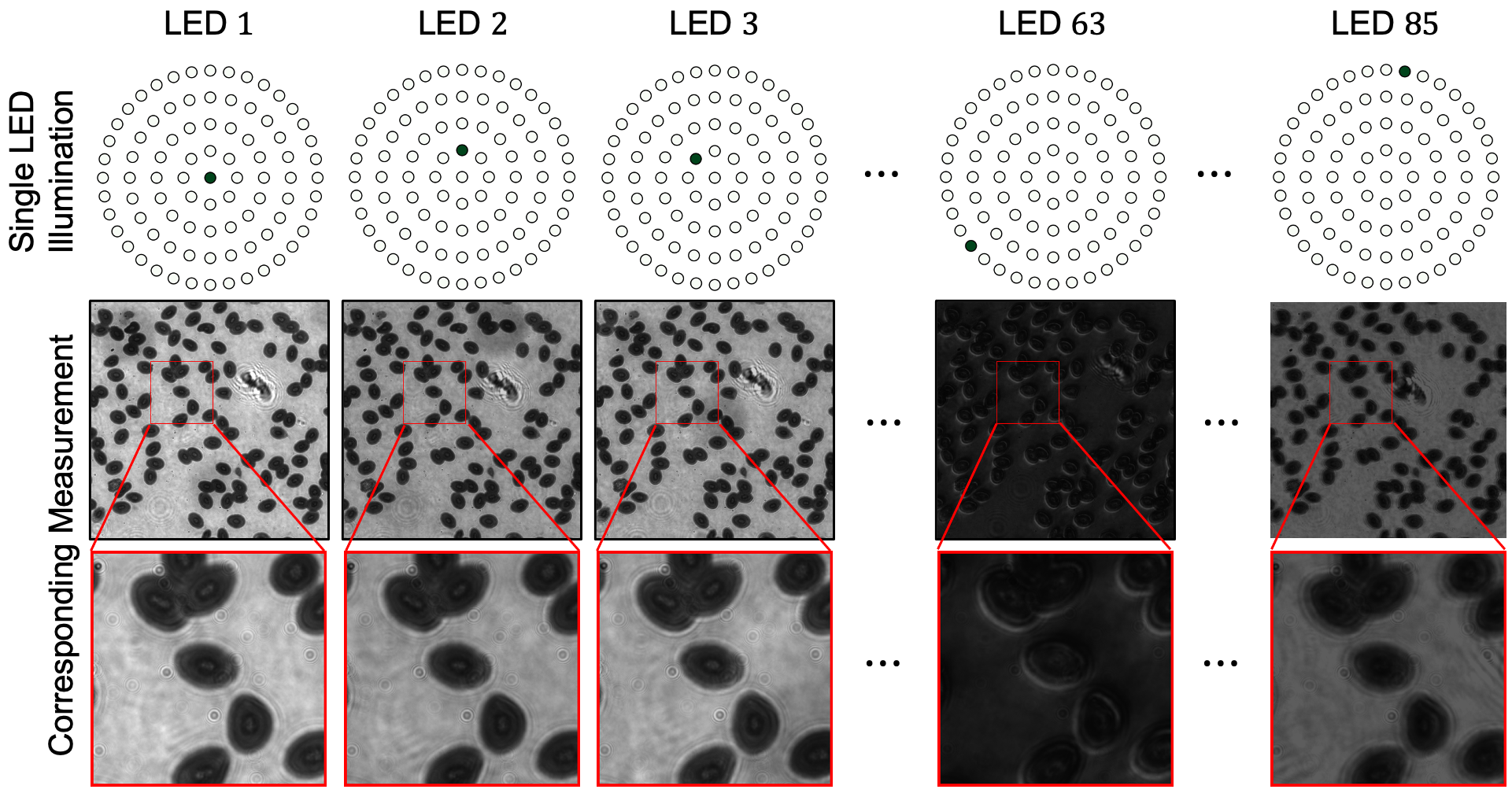} % second figure itself
        \caption{Sequential LED illumination.}
        \label{fig:dataset_real_singleLED}
    \end{minipage}
    \vspace{-1.5em}
\end{figure}

We collect an experimental dataset using an LED array microscope from slides of fixed frog blood smears (Eisco Labs), see Fig.~\ref{fig:dataset_real}. A programmable LED array of 257 LEDs in a concentric circular pattern, 6.5 mm average pitch (Spectral Coded Illumination), is mounted on an inverted microscope (Nikon Eclipse TE300). In this work, we utilize only the 85 centermost LEDs of the array. The distance $z$ from the LED array to the sample plane is 115 mm and the wavelength of light from the LED array is approximately 525 nm. The microscope objective is 40$\times$ magnification with a numerical aperture of 0.75 (Nikon CFI Plan Fluor). The image sensor captures 16-bit images with 2048 $\times$ 2048 pixels and a pixel size of 6.5 $\mu$m (pco.edge 4.2 LT). The dataset collected consists of 87 unique fields-of-view (i.e. objects). Each field-of-view was illuminated with a single randomly chosen LED pattern as follows: a 2.5 mm radius circle was randomly placed on the LED array and $\frac{1}{2}$ of the LEDs within that circle were randomly chosen to be illuminated. For reference, a dataset with sequential, single LED illumination was collected on a field-of-view, see Fig.~\ref{fig:dataset_real_singleLED}. This dataset is available for download at \url{https://doi.org/10.6084/m9.figshare.21232088.v1}.

\subsection{Results \& Discussion}

\begin{figure}[h]
  \begin{minipage}[b][][b]{.35\textwidth}
  
  \centering
  \vspace{0em}
  \begin{tabular}{| r | r | r|}
  
%Noise = 1e3
%Num Patterns = 1
%input_data = 'dataset_foam_v2_pac1'
%save_tag = 'foam_pac1'
%single pattern = False
%averaged over 10 examples

\hline
%P-VAE
\multicolumn{3}{|c|}{\textbf{P-VAE}, $n=1$}\\
\hline
   $m$ &   {\small Dirichlet} & {\small Deterministic}\\
\hline
                  1          &            -4.71 & -4.29\\
                  10        &            -4.12 & -3.19\\
                  100      &             0.31 & -2.27\\
                  1,000   &            18.82 & 2.13\\
                  10,000 &            18.43 & 1.97\\
\hline

\end{tabular}

\medskip

\begin{tabular}{| r | r |}
\hline
% Standard Iterative
\multicolumn{2}{|c|}{\textbf{Standard Iterative}}\\
\hline
   $n=29$ &   $n=1$ \\
\hline
   10.54 &  -3.96 \\

\hline
\end{tabular}

\captionof{table}{PSNR values as a function of dataset size $m$, averaged over 10 object examples.}
\vspace{-0.2em}
\label{table:foam_dataset_size_varypatt}
  
\end{minipage}\hfill
\begin{minipage}[b][][b]{.6\textwidth}
 \centering
\includegraphics[width=1.0\textwidth]{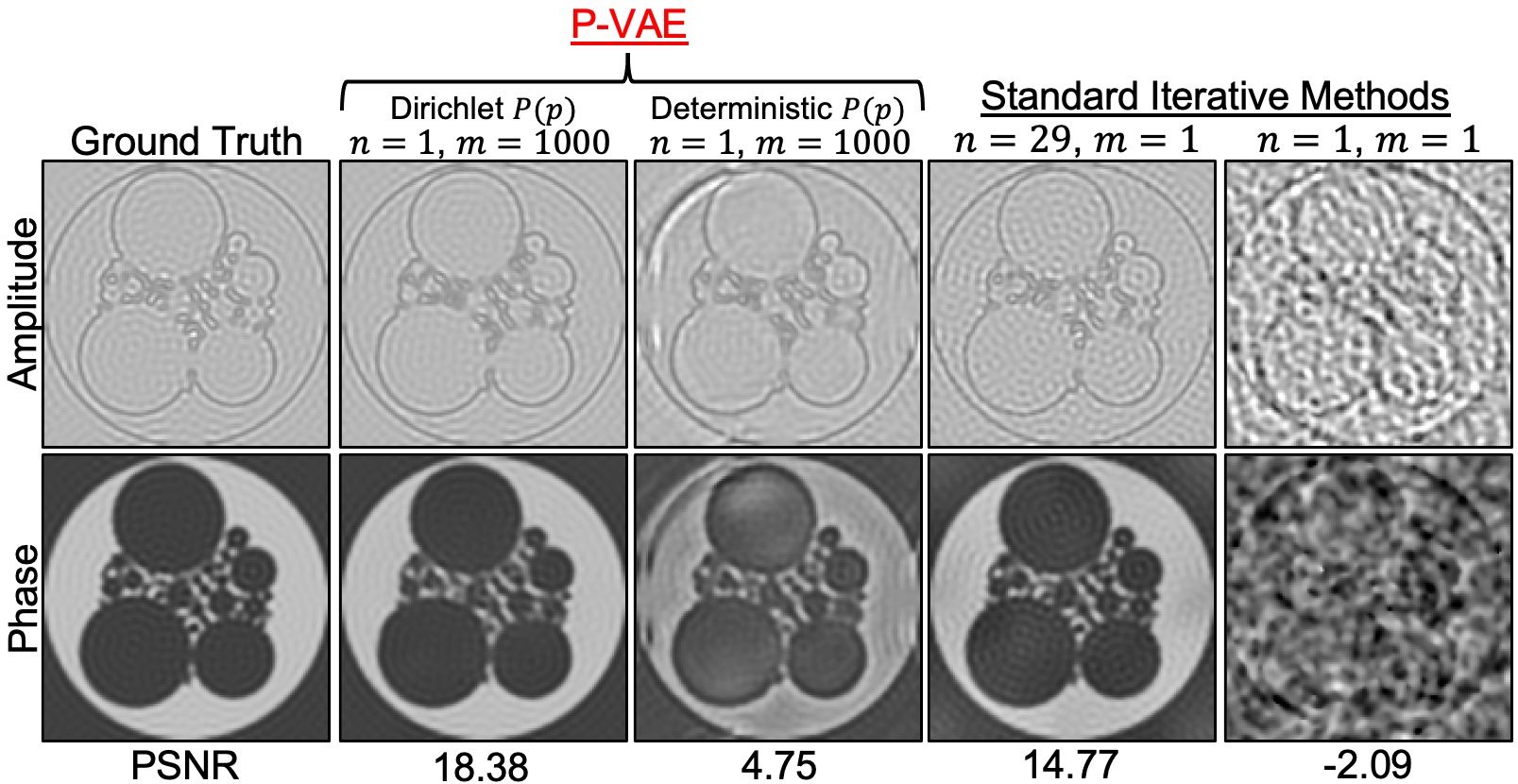}
\caption{Reconstruction results for a single object example. The results from the P-VAEs shown are with $m=1,000$. A constant $p$ over the dataset (deterministic $P(p)$) results in reconstruction artifacts.}
\label{fig:results_syn}
  \end{minipage}
\end{figure}

Code and documentation to reproduce our results are available at \url{https://github.com/vganapati/LED_PVAE}. We first train the physics-informed variational autoencoder (P-VAE) on the synthetic foam dataset, varying training dataset size. We compare a reconstruction sampled from the P-VAE, using one collected image (i.e.\ $n=1$), with conventional gradient-based iterative reconstruction using the same image (see \url{https://github.com/vganapati/LED_PVAE} for more details on implementation). We also evaluate iterative reconstruction with a stack of 29 single LED images (i.e.\ $n=29$). Table~\ref{table:foam_dataset_size_varypatt} compares the peak signal-to-noise ratio for the resulting reconstructions, averaged over 10 object examples, for both the cases of a Dirichlet $P(p)$ (different illumination patterns for every example) and a deterministic $P(p)$ (same illumination pattern for every example). Fig.~\ref{fig:results_syn} visualizes the comparison of peak signal-to-noise ratio (PSNR) among the methods for a single object. The results for standard iterative methods do not vary with dataset size. We note that the P-VAE with a Dirichlet $P(p)$ outperforms the standard iterative method with $n=29$, starting with a dataset size of $n=1,000$, and outperforms the standard iterative method with $n=1$, starting with a dataset size of $n=100$. A deterministic $P(p)$ does not perform as well, aligning with the intuition that diversity of measurements allows knowledge of the prior to be inferred.

We see that if the dataset size is large enough, and $P(p)$ is stochastic, learning from a dataset of limited measurements on different objects can lead to improvements of object reconstruction over conventional iterative methods. We emphasize the significance of our method in improving over standard iterative methods using less than 4\% of the data. We note that our procedure is similar to conventional object reconstruction through iterative optimization. The crucial difference is the joint reconstruction of multiple similar objects, with the pooling of knowledge between objects to find the prior distribution. We observe a slight decrease in performance of the P-VAE from a dataset size of $1,000$ to a larger dataset of $10,000$. The expectation is that a larger dataset would contain more knowledge of the prior distribution. However, we hypothesize that the limited capacity of the neural network may not allow for joint reconstruction on all the examples of a larger dataset. Future work aims to use hyperparameter optimization \cite{dumont_hyppo_2021} to investigate this hypothesis.

\begin{wraptable}{l}{0.375\textwidth}
\vspace{-1.5em}
  \begin{center}
% mnist, results as a function of number of patterns, averaged over 10 examples, for dataset 1e3, noise 1e3
% averaged over both slices
%input_data='dataset_MNIST_multislice_v2'
%save_tag='mnist'#'foam_pac1'
%single = False

\begin{tabular}{| r | r | r |}
\hline
   $n$ &   \textbf{P-VAE} &  \textbf{{\small Standard Iterative}} \\
\hline
                  1 &             7.55 &    0.07 \\
                  2 &            14.47 &  -0.02 \\
                  3 &            16.48 & 1.86 \\
                  4 &            18.61 &  3.69 \\
\hline
\end{tabular}

\smallskip

\begin{tabular}{| c  |}
\hline
% Standard Iterative
\textbf{Standard Iterative}, $n=29$\\

\hline
   21.17 \\
\hline
\end{tabular}

  \end{center}
%  \vspace{-.9em}
  \caption{PSNR values averaged for 10 objects from the 3D MNIST dataset.}
  \label{table:results_mnist}
  \vspace{-1.9em}
\end{wraptable}

\begin{figure}[h]
\centering
\includegraphics[width=1.0\textwidth]{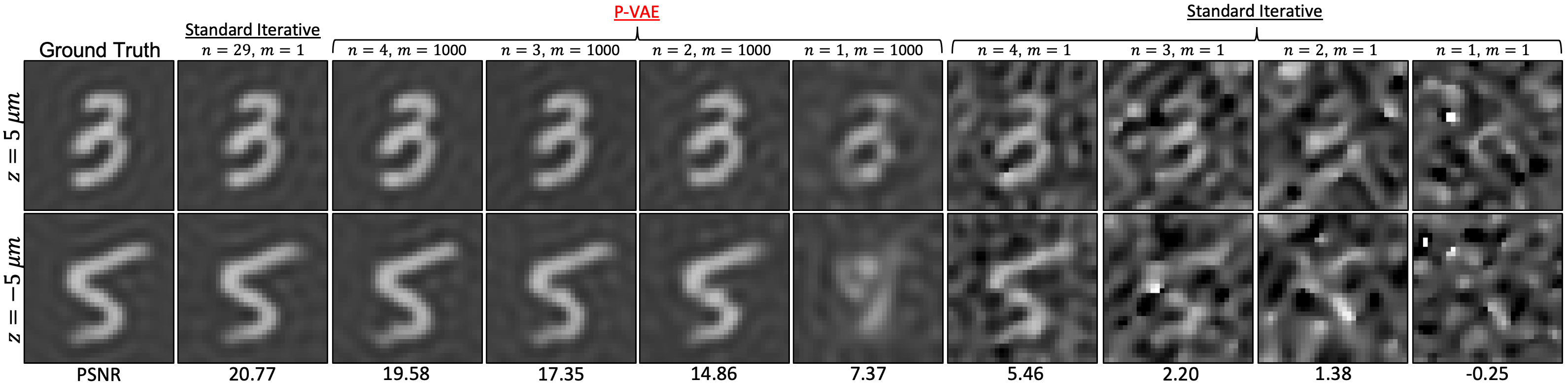}
\caption{Reconstruction results (phase) for an object of the 3D MNIST dataset.}
\label{fig:results_mnist}
\vspace{-1em}
\end{figure}

Results on the foam dataset demonstrate single-shot ($n=1$) imaging for a small array of 29 LEDs under the assumption of a thin, approximately 2D specimen. In the case of 3D imaging, more measurements may be needed. We train the P-VAE on the synthetic 3D MNIST dataset, with $1 \leq n \leq 4$, and show improvement for increasing number of measurements, see Table~\ref{table:results_mnist} and Fig.~\ref{fig:results_mnist}. We demonstrate improved reconstruction with the P-VAE over the standard iterative method using the same number of measurements $n$, though the the performance of the P-VAE doesn't exceed that of the standard method with $n=29$.

%\vspace{-0.5em}

%\begin{comment}

%\end{comment}

%\vspace{-1em}

\begin{wrapfigure}{r}{0.5\textwidth}
\vspace{-4em}
\centering
\includegraphics[width=0.5\textwidth]{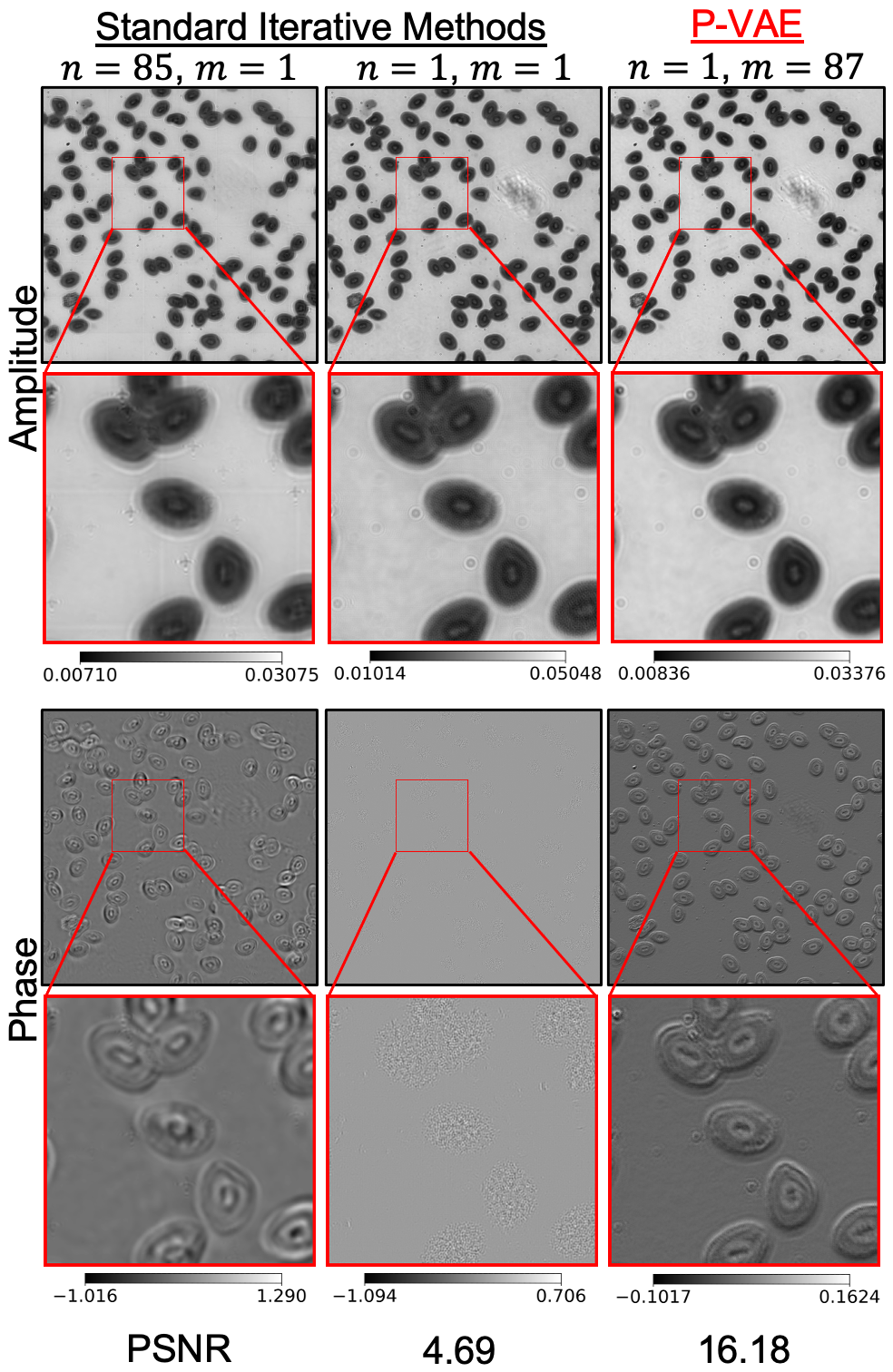}
\caption{Reconstruction results on the experimental dataset.}
\label{fig:results_frog}
\vspace{-4.0em}
\end{wrapfigure}

The P-VAE was tested on the experimental dataset and again compared against standard iterative methods. In the case of experimental data, there is no ground truth, so we utilize as reference the reconstruction obtained with a standard iterative method on a stack of 85 single LED images. We note that this is an imperfect metric, as we have seen the P-VAE outperform this method in certain cases of synthetic data. The experimental data reconstructions were conducted on an NVIDIA A100 GPU, and took 21 hours for reconstruction using a standard iterative method with $n=85$, 18 hours with standard iterative methods and $n=1$, and 47 hours for training the P-VAE. The standard iterative methods result in one object reconstruction, but the P-VAE results in 87 object reconstructions. The result from the P-VAE with $n=1$ outperforms the result obtained from the standard iterative method with $n=1$, when comparing to the reference, see Fig.~\ref{fig:results_frog}.

\section{Open Questions}

Though initial results are promising, more progress must be made to apply this imaging technique for scientific discovery.

\subsection{General Requirements}

In the case of real, experimental data, there is no ground truth to determine reconstruction accuracy (defined as average fidelity of samples of $P(O|M)$ to the true object by a metric such as mean-squared error, structural similarity, or peak signal-to-noise ratio). Correlation of these quantities to the training loss must be analyzed with synthetic data where the ground truth is known. Results must be shown to be robust, characterized by the worst-case loss of multiple training runs with the same hyperparameters. Results also should be stable, meaning small variance in loss among multiple training runs. Theoretical guarantees on the correctness of the calculated prior distribution need to be developed, given the form of the prior, the number of objects in the dataset, number of measurements per object, the distribution of parameters $P(p)$, and the forward model $P(M | O)$. Additionally, in the case of experimental data, the problem of model mismatch arises, where the specified likelihood $P(M | O)$ may have inaccuracies. The consequences of model mismatch need to be investigated in simulation with synthetic data.

\subsection{Choice of Measurement Parameters}

Each object of the dataset is measured with some parameters $p$. We aim for posterior distributions $P(O | M)$ with low entropy, for higher confidence in sampled reconstructions. The dependence of the posterior distribution entropy on the choice of $P(p)$ needs to be investigated. It is likely that the optimal choice of $P(p)$ depends on the prior distribution $P(O)$, which is unknown before measurement commences. The benefit of adaptively picking $p$, i.e.\ choosing $p$ for an object based on data collected from previous objects, can be evaluated in scenarios with ground truth or reference data.

\subsection{High Memory Management}

This work shows results from synthetic experiments using small-scale data. We demonstrate single-shot ($n=1$) and few-shot ($n \leq 4$) imaging for an array of 29 LEDs with measurements of $128 \times 128$ pixels or less. Results are also shown from real experimental data with single-shot imaging on an array of 85 LEDs with image size of $2048 \times 2048$ pixels. In scientific studies, the goal may be to achieve object reconstruction with larger arrays of around 800 LEDs, as more LEDs in the illumination pattern generally mean higher spatial resolution \cite{zheng_wide-field_2013}. For these large-scale, real-world problems, the number of measurements $n$ required with this framework may be $\gg 1$. For other computational imaging modalities, such as computed tomography, thousands of measurements may be collected per object \cite{marchesini_sparse_2020}. A single set of measurements on an object in x-ray holographic nano-tomography can be on the order of 100s of gigabytes \cite{kuan_dense_2020}. Solving jointly for the posteriors of an object set will require techniques for high memory management.

\subsection{Video Reconstruction}

Fast acquisition times are critical to imaging dynamic objects, to avoid motion blur and instead capture movement and changes. Like other computational imaging systems, the LED array microscope allows for increased functionality at the cost of temporal resolution. For many biological applications, imaging live, changing specimens is of key importance to scientific discovery. The outlined framework can be utilized for video reconstruction by treating each frame as a separate reconstruction problem. In particular, snapshots of the same object at different timestamps $t=1,2,...,m$ can be thought of as the objects of the training dataset  $\{O_1, O_2, ..., O_m\}$. However, this assumes that snapshots are independent and does not take into account rich temporal dependencies among frames that can lead to improved reconstruction and elimination of ``jumpy'' artifacts \cite{baraniuk_compressive_2017, schlemper_deep_2018, mur_recurrent_2020}. Approaches with recurrent neural network architectures may allow for improved video reconstruction.

\section{Conclusion}

We present a novel self-supervised framework, based on the mathematics of variational autoencoders, for reconstruction in computational imaging. The framework is prototyped for improving the temporal resolution of LED array microscopy, and validated with synthetic and experimental data. As the framework is self-supervised and does not require ground truth or reference reconstructions for training, it is amenable for use in applications of scientific discovery, where both temporal and spatial resolution are needed. We outline future directions of research, releasing our code at \url{https://github.com/vganapati/LED_PVAE} and our experimental data at \url{https://doi.org/10.6084/m9.figshare.21232088.v1} for further development.

\begin{ack}

This work was supported in part by the U.S. Department of Energy, Office of Science, Office of Workforce Development for Teachers and Scientists (WDTS) under the Visiting Faculty Program (VFP) and the AAUW Research Publication Grant in Engineering, Medicine and Science. The authors thank Srinivas Turaga and Jan Funke at Janelia Research Campus and Vincent Dumont at Lawrence Berkeley National Laboratory for participating in helpful discussions.

\end{ack}

\bibliography{main}

%%%%%%%%%%%%%%%%%%%%%%%%%%%%%%%%%%%%%%%%%%%%%%%%%%%%%%%%%%%%

\end{document}